\documentclass[11pt,letter, notitlepage]{article}
\usepackage{graphicx}
\usepackage[letterpaper]{geometry}
\usepackage{authblk}
\usepackage{url}
\usepackage{dcolumn}
\usepackage{upgreek}
\usepackage{epstopdf}
\newcommand{\Gres}{\ensuremath 6.674\,252(122)\times 10^{-11}}

\newcommand{\Gunit}{\ensuremath \mbox{m}^3\,\mbox{kg}^{-1}\,\mbox{s}^{-2}}
\newcommand{\Lmax}{\ensuremath L_\mathrm{max}}
\renewcommand{\mu}{{\ensuremath \upmu}}

\title{Reflections on a Measurement of the Gravitational Constant Using a Beam Balance
and 13 Tons of Mercury}
\author{%%%% Author details
S. Schlamminger$^{1}$, R.E.~Pixley$^{2}$, F.~Nolting$^3$, J.~Schurr$^4$ and  U.~Straumann$^{2}$}

\affil{$^{1}$National Institute of Standards and Technology, Gaithersburg, MD, USA.\\
$^{2}$Physik-Institut der Universit\"at Z\"urich, Z\"urich, Switzerland.\\
$^{3}$Paul Scherrer Institut, Villigen, Switzerland. \\
$^{4}$Physikalisch Technische Bundesanstalt, Braunschweig, Germany.
}

\begin{document}
%\thanks{Email: stephan.schlamminger@nist.gov}
\maketitle
\begin{abstract}
In 2006, a final result of a measurement of the gravitational constant $G$ performed by researchers
at the University of Z\"urich was published. A value of
$G=\Gres\,\Gunit$ was obtained after an experimental effort that lasted over one decade. Here,
we briefly summarize the measurement and discuss the
strengths and weaknesses of this approach.
\end{abstract}
%%%%%%%%%%%%%%%%%%%%%%%%%%%

%%%%%%%%%% Insert the texts which can accomdate on firstpage in the tag "fmtext" %%%%%

%\begin{fmtext}
\section{Introduction}

The existence of the Z\"{u}rich $G$ experiment is due to an article published in 1986 by Fischbach et al. ~\cite{Fischbach86} analyzing old data taken by
von E\"otv\"os~\cite{eot22} to test the universality of free fall. A deviation was found in the intermediate-range (about 200\,m) coupling, giving rise
to the so-called fifth force. The existence of such a fifth force at the strength conjectured in~\cite{Fischbach86} was quickly found to be in
error~\cite{Stubbs87}. However, Fischbach's article started a renaissance of gravity experiments at universities world wide. Walter K\"undig at the
University of Z\"{u}rich started an experiment aimed at measuring the gravitational attraction between water in a storage lake (Gigerwald lake) and two
masses suspended from a balance~\cite{Cornaz94}. The experiment was conducted in two different configurations. First, the test masses were vertically
separated by 63\,m, later by 104\,m. The magnitude of the gravitational attraction was measured as the water level varied seasonally over the course of
several years. As a result of the experiment, two measurements of $G$ for an interaction range of 88 m and 112 m were obtained with relative
standard-uncertainties of $1\times 10^{-3}$ and $7\times 10^{-4}$, respectively~\cite{Hubler95}. The results were consistent with each other and consistent
with the value of $G$ from laboratory determinations. Even today, 20 years after this experiment, its results place the most stringent limit on a possible
violation of the inverse square law at distances ranging from 10\,m to 100\,m.   The largest contribution to the uncertainty of each result came from the
ambiguous mass distribution of the lake.  It was unclear how far the lake water penetrated the shore composed mostly of scree. It was immediately
recognized that one could use the same method for a precise determination of the gravitational constant if only one had a better defined lake.

From this line of thinking, the concept of measuring $G$ in the laboratory was conceived  and the design of the experiment started in 1994. Conceptually
the experiment is similar to the Gigerwald experiment with one difference: the ''lake'' was confined to two well characterized stainless-steel vessels each
holding 500\,L of liquid. In a first experiment, water was used; later, the water was replaced with mercury yielding a much larger signal.

The Z\"{u}rich big $G$ experiment ended officially in 2006,  when a final report~\cite{Schlamminger06} on the experiment was published. The relevant
details of the experiment have been summarized in the final report, two theses~\cite{Nolting98, Schlamminger02} and several shorter
reports~\cite{Schurr98a, Schurr98b, Nolting99a,Nolting99b,Nolting00,Schlamminger02b}.

\section{The experiment's principle}
\begin{figure}[htb]
\begin{center}
\includegraphics[width=5.25in]{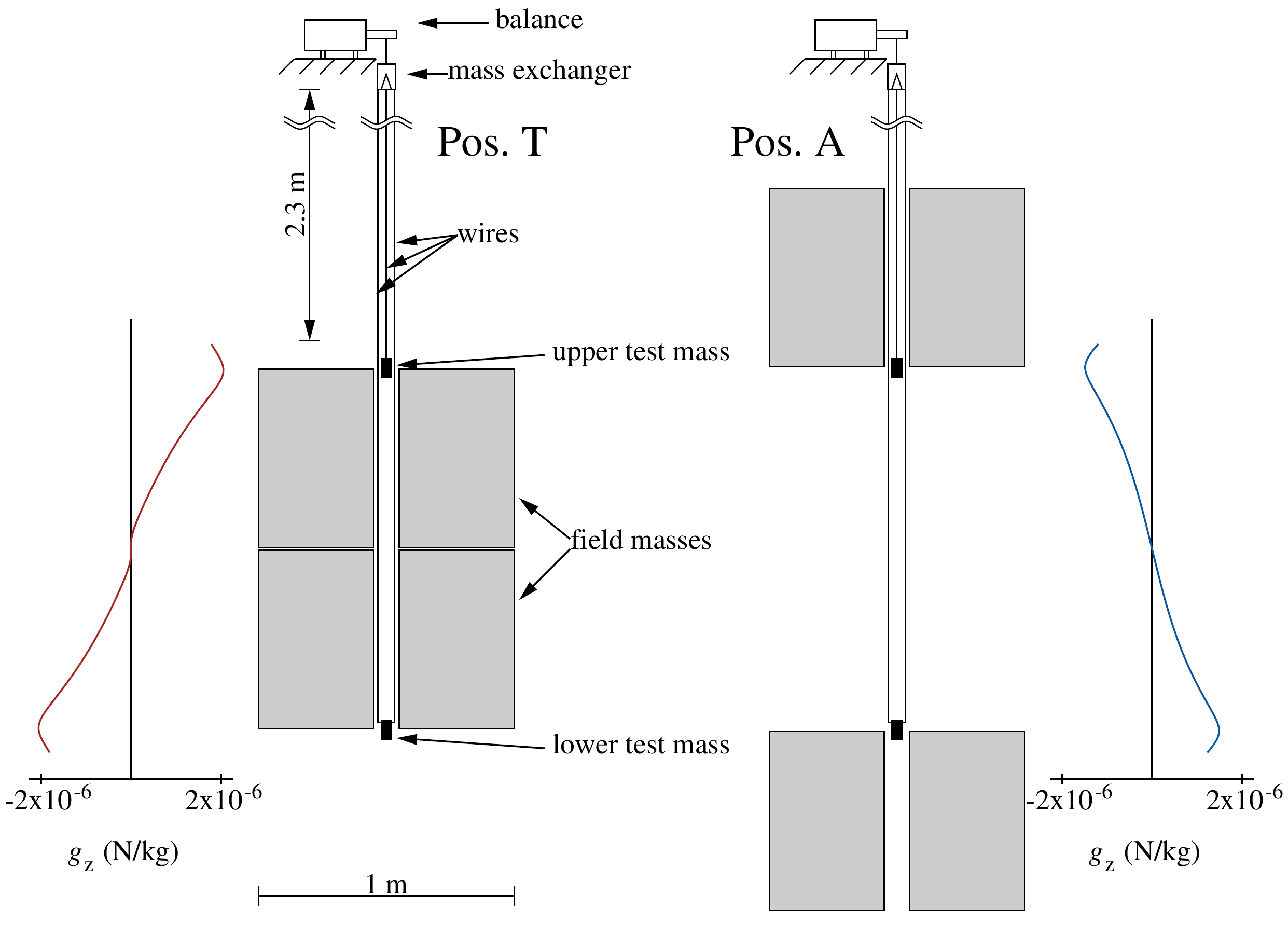}
\end{center}
\caption{Principle of the Z\"{u}rich $G$ experiment. Either of the two test masses is suspended from the
balance. The field masses are either in the position T (together) or A (apart) as shown on the left and
 right side of the figure respectively. The graphs to the side of the field masses show the vertical part
 of the gravitational field generated by the field masses along the symmetry axis in the center of the
 hollow cylinders. A downward force corresponds to a positive sign.}
\label{fig:principle}
\end{figure}

The principle of the experiment is shown in figure~\ref{fig:principle}. A gravitational field is generated by two large cylinders labeled field masses
(FMs). The gravitational field can be modulated by moving the FMs. During measurement, the FMs are in either one of two positions, labeled T for together
and A for apart. Two test masses (TMs) are used to probe the  gravitational field. The two test masses are alternately, but never concurrently, connected
to a mass comparator (balance) at the top of the experiment each by a set of two wires and a mass exchanger. While the mass comparator is
calibrated in units of kg, it is used as an instrument to measure vertical force with high accuracy. The balance is calibrated
 by adding calibration masses to the mass pan. The reading of the balance can be converted into
 a force by multiplying with the value of the local acceleration,
$g=9.807\,233\,5(6)\,$m/s$^2$ which was  measured at the site.

In the together/apart state, the difference in the force on the upper and lower test mass are given by
\begin{eqnarray}
\Delta F_\mathrm{T} &=& m_\mathrm{u} g(z_\mathrm{u}) + F_\mathrm{z}(\mathrm{T},\mathrm{u})
                    -   m_\mathrm{l} g(z_\mathrm{l}) - F_\mathrm{z}(\mathrm{T},\mathrm{l}) \\
\Delta F_\mathrm{A} &=&  m_\mathrm{u} g(z_\mathrm{u}) + F_\mathrm{z}(\mathrm{A},\mathrm{u})
                    -   m_\mathrm{l} g(z_\mathrm{l}) - F_\mathrm{z}(\mathrm{A},\mathrm{l}),
\end{eqnarray}
where $F_\mathrm{z}(\mathrm{A}/\mathrm{T},\mathrm{u}/\mathrm{l})$ denotes the vertical gravitational force between the complete field mass assembly in
the A/T position and the upper/lower test mass. The difference, $\Delta F_\mathrm{T}-\Delta F_\mathrm{A}$ is the
 second difference or gravitational signal and is given by
\begin{equation}
s = \Delta F_\mathrm{T}- \Delta F_\mathrm{A} =
F_\mathrm{z}(\mathrm{T},\mathrm{u})-F_\mathrm{z}(\mathrm{A},\mathrm{u})-F_\mathrm{z}(\mathrm{T},\mathrm{l})+F_\mathrm{z}(\mathrm{A},\mathrm{l}) = G \Gamma.
\end{equation}
It can be seen that the mass difference between the upper and lower test masses, as well as the value of local gravity and its gradient on the test masses
vanish. In our convention a positive force points downward (increasing balance reading). With this convention, $F_\mathrm{z}(\mathrm{T},\mathrm{u})$ and $F_\mathrm{z}(\mathrm{A},\mathrm{l})$
are positive, but $F_\mathrm{z}(\mathrm{T},\mathrm{l})$ and $F_\mathrm{z}(\mathrm{A},\mathrm{u})$ are negative. The second difference can be written as a product of $G$ and a mass integration
constant $\Gamma$. The mass integration constant has units of kg$^2$/m$^2$ and depends solely on the mass distributions within the field masses, the test
masses, and their relative positions in the two states.

The size of the gravitational signal $s/g \approx 785\,\mu$g was determined with an accuracy of 14.3\,ng using a modified commercial mass comparator
(Mettler Toledo AT1006\footnote{Certain commercial equipment, instruments, or materials are identified in this paper in order to specify the experimental
procedure adequately. Such identification is not intended to imply recommendation or endorsement by the National Institute of Standards and Technology, nor
is it intended to imply that the materials or equipment identified are necessarily the best available for the purpose}). Off the shelf, this type of mass
comparator is used at national metrology institutes to compare 1~kg weights with each other using the substitution method. The mass comparator is
essentially a sophisticated beam balance, where a large fraction of the load on the mass pan is compensated by a fixed counter mass. Only a small part of
the gravitational force on the mass pan is compensated by an electromagnetic actuator consisting of a stationary permanent magnet system and a current
carrying coil attached to the balance beam. The current in the coil is controlled such that the beam remains in a constant position. The coil current is
precisely measured and converted into kg.  This value is shown on the display and can be transferred to a computer. Typically, the dynamic range of such a
comparator is 24\,g around 1000\,g with a resolution of 100\,ng.  The comparator used in the present experiment was modified in two ways: First, the mass
comparator was made vacuum compatible by stripping it of all plastic parts and separating the electronics from the weighing cell. The electronics outside
the vacuum vessel was connected to the balance via vacuum feed throughs. Second, the dynamic range was reduced to 4\,g thereby decreasing the
resolution to 16.7\,ng by simply reducing the number of turns of the coil by a factor of six. Hence, for the same applied current to the coil, the force
was reduced by a factor six. Since the coil current was measured with the same electronics, the mass resolution decreased by a factor six. An additional
decrease to 12.5\,ng was achieved by modifying software in the balance controller.

\begin{figure}[ht!]
\begin{center}
\includegraphics[width=5.25in]{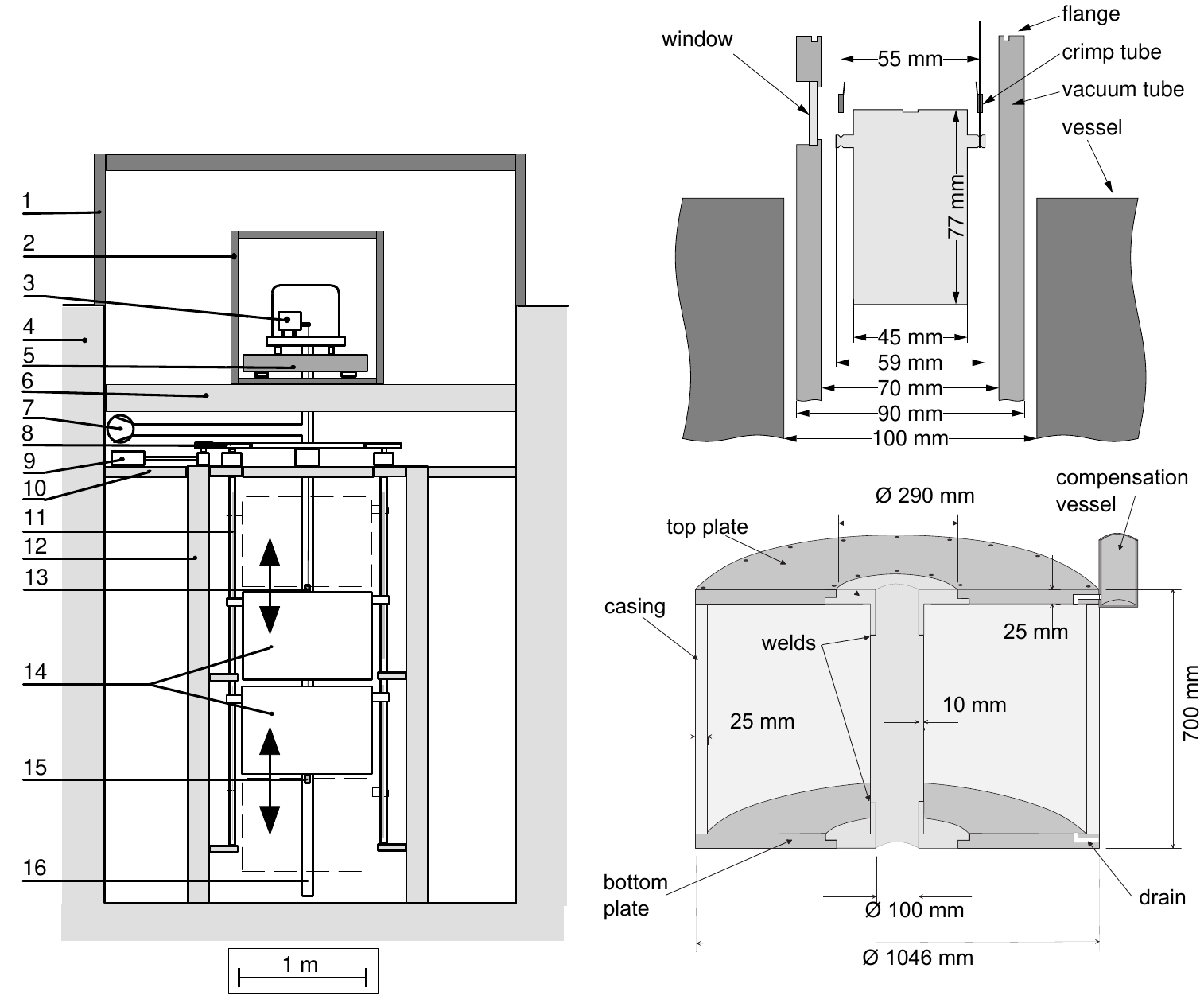}
\end{center}
\caption{ On the left is a side view of the experiment. Legend: 1=measuring room enclosure,
 2=thermally insulated chamber, 3=balance, 4=concrete walls of
the pit, 5=granite plate, 6=steel girder, 7=vacuum pumps, 8=gear drive,
 9=motor, 10=working platform, 11=spindle, 12=steel girder of the main support,
 13=upper TM, 14=FM's, 15=lower TM, 16=vacuum tube. On the lower right,
 a drawing of one of the vessels holding the liquid mercury.  On the upper right, a cross-sectional view of the lower test mass in the vacuum tube.}
 \label{fig:tank:setup}
\end{figure}

The field masses are cylindrical vessels made from stainless steel, each with an inner volume of 500\,L. The lower right panel of
figure~\ref{fig:tank:setup} shows a drawing of one vessel. Each vessel was filled with 6760 kg of mercury. A liquid was \ used to ensure a homogeneous
density and mercury was chosen due to its high density of 13.54\,g/cm$^3$. The gravitational signal produced by the liquid is proportional to its density.
Therefore, the gravitational signal due to the mercury is 13.5 times larger than that of water. However, since the contribution of the stainless steel
vessels needs to be taken into account, the mercury filled vessels produce only 4.1 times the signal of the water filled vessels. The vessels contribute
about $60\,\mu$g to the signal, which is about 7.6\% of the signal obtained with the mercury filled vessels, but 55\%  of the signal obtained with water
filled vessels.

The vessels were evacuated prior to filling them with mercury to avoid trapping air. The mercury was delivered in 395 flasks, each weighing about 36.5\,kg (34.5\,kg mercury and 2\, kg due to the steel flask). Each flask was weighed before and after its content was transferred to one of the two vessels using an evacuated transfer system. This painstakingly careful work led to a relative standard uncertainty of the mercury mass of $1.8\times 10^{-6}$ and $2.2\times 10^{-6}$ for the upper and lower vessel, respectively. The mercury for this experiment was leased, i.e., after the experiment was dismantled in December 2002 the mercury was sent back to the supplier in Spain.

The upper right panel in figure~\ref{fig:tank:setup} shows a cross sectional drawing of the lower test mass. The test masses were made from oxygen free,
high-conductivity (OFHC) copper and were coated with a thin layer of gold to prevent oxidation; no ferromagnetic adhesion layer was used in this coating
process. Copper was the material of choice to avoid magnetic forces on the masses. Each test mass was a few grams less than 1100\,g.  The test masses were
100\,g heavier than the nominal 1\,kg load of the balance. This was possible, because 100\,g was removed from the mass pan of the mass comparator to make
it vacuum compatible. The stability of the two gold plated copper masses was acceptable during the course of the experiment. Over the five years of usage,
the value of each mass was determined eight times at a calibration laboratory. The mass of each test mass varied by less than $440\,\mu$g or in relative terms less than $4\times10^{-7}$.

The mass exchanger allows either one of the two test masses to be connected to the balance. Each test mass is suspended by two wires from an aluminum cross
bar. The two cross bars are perpendicular to each other and vertically displaced. Each cross bar can either be suspended from a stirrup that is suspended from the 
mass pan of the balance or to a hydraulic actuated arm. Each arm can be moved vertically by a few millimeters. Lowering the arm places the crossbar onto
the stirrup and hence connects the test mass to the balance. The mass exchange algorithm was programmed such that there was always a 1100\, g load on the
balance. During mass exchange one arm was lowered, while the other was simultaneously raised.

To avoid buoyancy and other gas pressure forces, the vessel containing the balance and a tube surrounding the test masses were evacuated to pressures
around $10^{-4}\,$Pa. The upper right panel in figure~\ref{fig:tank:setup} shows a drawing of the lower test mass, the vacuum tube surrounding it and its position relative to the lower 
field mass in the apart position.

A view of the experiment described above is given in the left panel of figure~\ref{fig:tank:setup}. The experiment was located in a pit at the Paul
Scherrer Institute (PSI) in Villigen, Switzerland. The pit was divided into an upper and a lower room separated by a false floor. The lower room contained
the TM's and FM's. The upper room contained a thermally insulated chamber housing the vacuum vessel with the balance. The vacuum vessel rested on a granite
plate that sat on two steel girders spanning the pit. This mounting ensured that the balance was decoupled from the field mass assembly which was anchored
to the bottom of the pit. The upper room of the pit housed the control electronics, the computers and the vacuum pumps. The temperature inside the thermally insulated chamber surrounding the balance was actively controlled; the temperature was stable within 0.01\,K. The air temperature of the lower part of the pit was actively controlled to within $\approx0.1\,$K. Thirteen tons of mercury make a great thermometer. A temperature rise of only a few kelvin would have been enough to bring the mercury level to the top rim of the compensation vessel.

\section{Linearity and calibration of the mass comparator}

\label{sec:cal}

A large amount of time was spent understanding and improving the calibration procedure of the mass comparator that was employed in the first  measurement
\cite{Schurr98a} of $G$ made with this apparatus.  However, questions remained about the linearity of the balance, i.e., is the calibration of the mass comparator that was performed with a 1\,g mass valid at a signal level of 785\,$\mu$g (see top panel of Fig. 3)? Nonlinearity measurements on a laboratory balance (AT261) with a measuring range of 200\,g were performed by engineers at Mettler Toldeo. By scaling their results to the mass comparator used in the $G$ experiment, an upper limit for a possible bias due to the nonlinearity was estimated to be $200\times 10^{-6}$ of the result. In the 1998 result, the nonlinearity of the balance was the largest entry in the uncertainty budget.

Calibrating a balance in the range of 800\,$\mu$g is not a simple problem. The obvious solution of using a calibration mass of approximately 1\,mg
does not work for the present $G$ measurement as the uncertainty of such a small mass relative to the international prototype of the kilogram is of the
order of $500\times 10^{-6}$, i.e., more than an order of magnitude  larger than the desired accuracy of the $G$ measurement. E.~Holzschuh suggested the
principle that was finally used to solve this problem. The basic idea is the following: The nonlinearity of a balance can be averaged away by measuring the
gravitational signal, not just at one point of the transfer function, but at many  points equally spaced within a calibration range defined by a
calibration mass having a sufficiently accurate absolute mass. A deeper rational for this method comes from the fundamental theorem of analysis which
basically says that the average of the local slopes of a Riemann integrable function is the same as the slope of a line connecting the start and the end
points of the averaging interval. The different points of the measurement are easily obtained by adding small masses to the balance pan having roughly
equal weights. Their masses need to be known only relative to one another. For the implementation of this calibration method, we employed 256 mass steps of
approximately 785\,$\mu$g over a calibration range of 200\,mg.
%The absolute mass of the two 0.1\,g calibration masses employed was each
%known to 3.5\,$10^{-7}$g.

Although the basic principle of this method is based on equal mass steps over a calibration range, it is difficult to make small masses with exactly equal
mass. In addition, it was not always possible to make all mass steps due to malfunction of the auxiliary mass handler. A somewhat more general analysis method was
developed to overcome these problems. For this purpose, a series of Legendre polynomials with an arbitrary highest order $L_\mathrm{max}$ was chosen.
Hence, the calibrated reading of the balance $f(u)$ as a function of the mass $u$ on the mass pan can be written as
\begin{equation}
  f(u)=\sum_{\ell=0}^{L_\mathrm{max}} a_\ell P_{\ell} (\xi),
\end{equation}
where $P_{\ell}$ is a Legendre polynomial of order $\ell$ and $\xi=2u/u_\mathrm{max}-1$. Two constraints, $f(0)=0$ and $f(C)=C$, reduce the number of
degrees of freedom  from $L_\mathrm{max}+1$ to $L_\mathrm{max}-1$   where $C$ is the sum of the two calibration masses. Thus, the values of $a_0$ and $a_1$ are given by
\begin{equation}
  a_0=C/2-\sum_{\mathrm{even} \; \ell=2}^{L_\mathrm{max}} a_\ell
\;\;\;\mbox{and}\;\;\;
  a_1=C/2-\sum_{\mathrm{odd} \; \ell=3}^{L_\mathrm{max}} a_\ell.
\end{equation}
The remaining $L_\mathrm{max}-1$ parameters $a_\ell$, and the size of the signal $s/g$, were obtained by minimizing
\begin{displaymath}
   \chi^2=\sum_{n=1}^N \left [f(u_n+s/g)
   -f(u_n)-y_n\right]^2 \sigma_n^{-2},
\end{displaymath}
where $y_n$ is the calibrated reading obtained with offset $u_n$ and $\sigma_n$ is the statistical standard deviation of the reading. The minimization of
$\chi^2$ is straightforward. Since $s/g$ is the only nonlinear parameter, a one parameter search for a minimum is all that is required. The $a_\ell$
parameters are linear and can be solved by a trivial matrix inversion. The number of parameters required for a reasonable fit can be determined from the
value  of $\chi^2$.

The small masses used for shifting the measuring  point are referred to as auxiliary masses (AM). Two sets of AMs were employed. One set (AM-1) contained
16 masses with an average value of $783\,\mu$g and a standard deviation of $1.5\,\mu$g. The other set (AM-2) contained 16 masses with an average mass of 16
times $783\,\mu$g and a standard deviation of $2.3\,\mu$g. With combinations of these two sets of masses it is possible to have 256 approximately equal
mass steps in the calibration interval from 0 to 200\,mg . The masses of AM-1 and AM-2 were made from stainless steel wire with a diameter of 0.1\,mm and
0.3\,mm, respectively.

\begin{figure}[ht!]
\begin{center}
\includegraphics[width=5.25in]{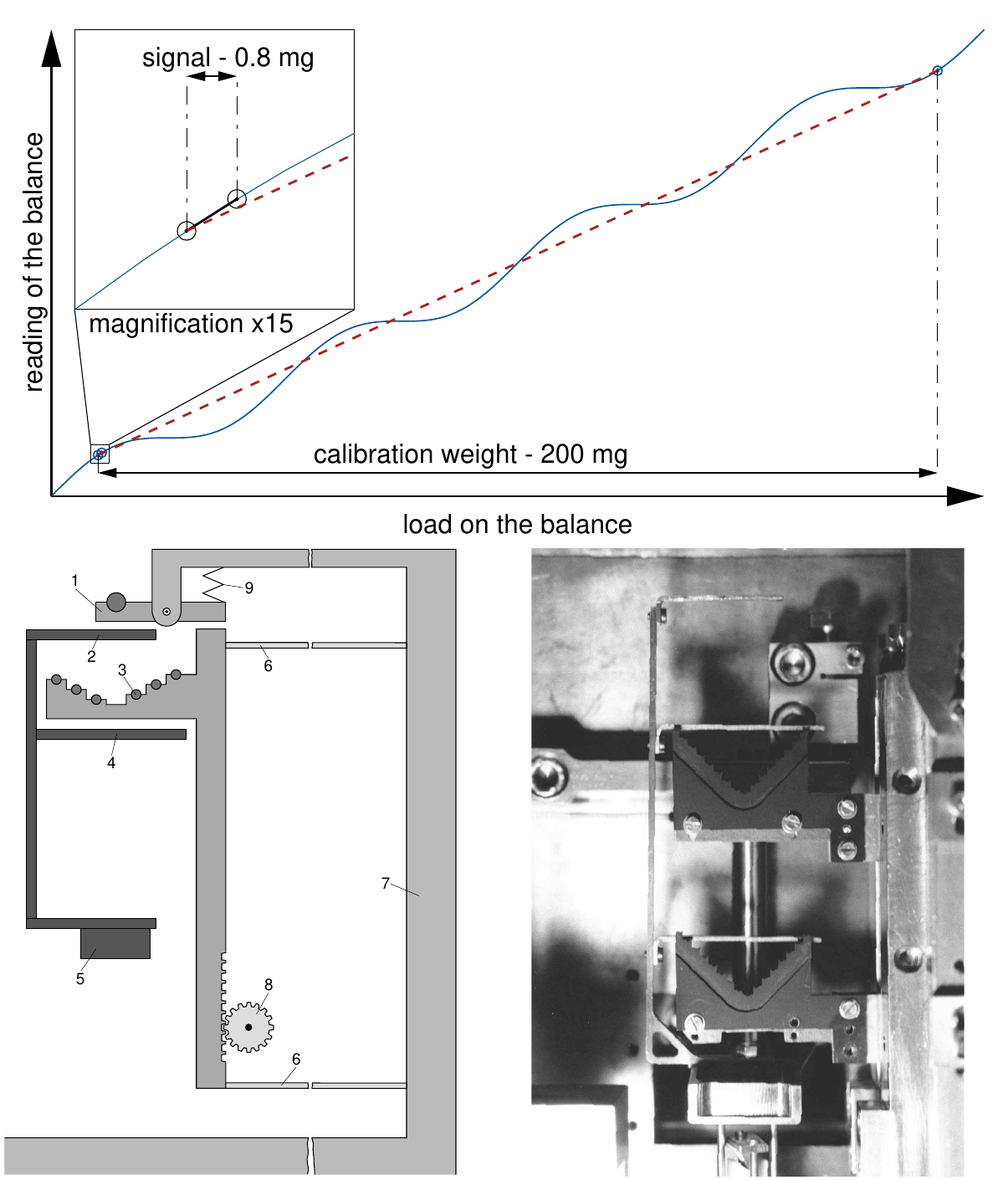}
\end{center}
\caption{The upper panel schematically illustrates the effect of a nonlinearity in the balance on the measurement. The top graph shows the transfer function of the
balance, i.e. reading on the vertical axis as a function of load on the horizontal axis. Since the calibration weight is over 250 times larger than the
signal it is not clear that the slope of transfer function at the working point is identical to that of the calibration. The figure in the lower left panel
is a schematic drawing of the mass handler with the following legend: 1=pivoted lever pair holding a CM, 2=narrow strip to receive the CM, 3=double stair
case pair holding AM's, 4=narrow strip to receive AM's, 5=balance pan, 6=flexure strip,
 7=frame, 8=rack and pinion, and 9=coil spring. The lower right panel
shows a photograph of the mass handler during installation.} \label{fig:nonlin}
\end{figure}

The device for loading the masses on the balance is shown in the lower left panel of figure~\ref{fig:nonlin}. Initially, each set of masses rests on one of
the two separately controlled double staircases bracketing a narrow metal strip attached to the balance pan. The steps are 2\,mm high  and  2\,mm wide.
They are arranged in the form of a ``V''. The vertical position of each staircase is controlled by a rack and pinion device driven by a stepper motor.
Lowering the staircase places alternately a mass from the right side of the ``V'' and then one from the left side on the balance pan. This keeps the
off-center loading of the balance pan small.

Very little nonlinearity was observed. The value of $s/g$
obtained from the $\chi^2$ minimization above changed from $784.899\,4\,\mu$g for a linear transfer function, i.e., $L_\mathrm{max}=1$ to a value of
$784.900\,5\,\mu$g for $59\le\Lmax\le66$. For the final published result the latter value was used, yielding a value for $G$ that is only $1.4\times
10^{-6}$ larger than the value we would have obtained with no correction. The numbers above are a reflection of the good design of the balance and provide increased confidence in the measurement result, since the value of $G$ is independent of the subtleties of the data analysis. In the end, it was worth tracking down this problem and putting the specter of nonlinearity in this experiment to rest, once and for all.

The mass handler also doubles as a device to place either of two calibration masses (CMs) on the balance pan. Each rested on a spring-loaded double lever.
By raising each staircase, one of the CMs is placed on the balance pan. Each calibration mass had a nominal value of 100\,mg and was made from stainless
steel wire.

The  CMs were calibrated at the Swiss Federal Institute of Metrology (METAS) before and after each measurement campaign.  In total, three measurement
campaigns were performed. Unfortunately in the last two campaigns problems with the mass handler rendered the data unsuitable for a $G$ determination.
Hence, only data from the first campaign were used to determine the final value of $G$.

Water adsorption on the calibration masses is a concern. The three calibrations made by METAS were performed in air, i.e., with an ubiquitous water film on
the steel wires. In the experiment, however, the weight in vacuum was important. To experimentally investigate this effect, wires with two different
diameters, 0.96\,mm and 0.5\, mm, were used to make the CMs.  Hence, the surface area of the CM with the thinner wire was almost twice that of the thicker
wire. It was found that the weight difference of the two CMs in vacuum was consistent with their difference measured by METAS in air. Hence, no sorption
effect could be detected. We placed an upper limit on the sorption effect using sorption coefficients from the literature \cite{Schwartz94}. Since the
vacuum environments are difficult to compare, a generous relative standard uncertainty of 100\% was assigned to these values.

Having two calibration masses made the calibration process more robust. The mass difference between the  CMs  determined in the first and second
calibrations made by METAS changed by $1.0\pm 0.5\,\mu$g, i.e., $5\pm 2.5\times 10^{-6}$ of their combined weight. Their weight difference in vacuum was
found to be consistent with the first METAS calibration. Since in the second and third measurements in vacuum, the CM weight difference was consistent
with the second and third METAS calibration, only one viable hypothesis could explain these observations: One wire lost a small dust particle or even a
part of the wire itself during transfer from the measurement at PSI to the calibration measurement at METAS. Note that the wire that lost the particle was the
thicker wire that had jagged edges from being cut by a wire cutter. Based on this hypothesis we used the values obtained by METAS during the first calibration to analyze our data. 

From  these measurements, we are confident that our result for G is based on an SI traceable calibration of the measured gravitational signal. In the end,
a relative standard uncertainty of $7.3\times 10^{-6}$ was assigned to calibration and nonlinearity. We acknowledge that the observed mass change of
$1\,\mu$g in one of the calibration masses was not desirable. However, after careful review of the mass differences obtained in air and in vacuum we have
found the only possible explanation. The nonlinearity of the balance has a small effect on the result and was certainly not as large as the conservative uncertainty given in 1998.

\section{Measurements and Result}\label{sec:result}
\begin{figure}[ht]
\begin{center}
\includegraphics[width=2.6in]{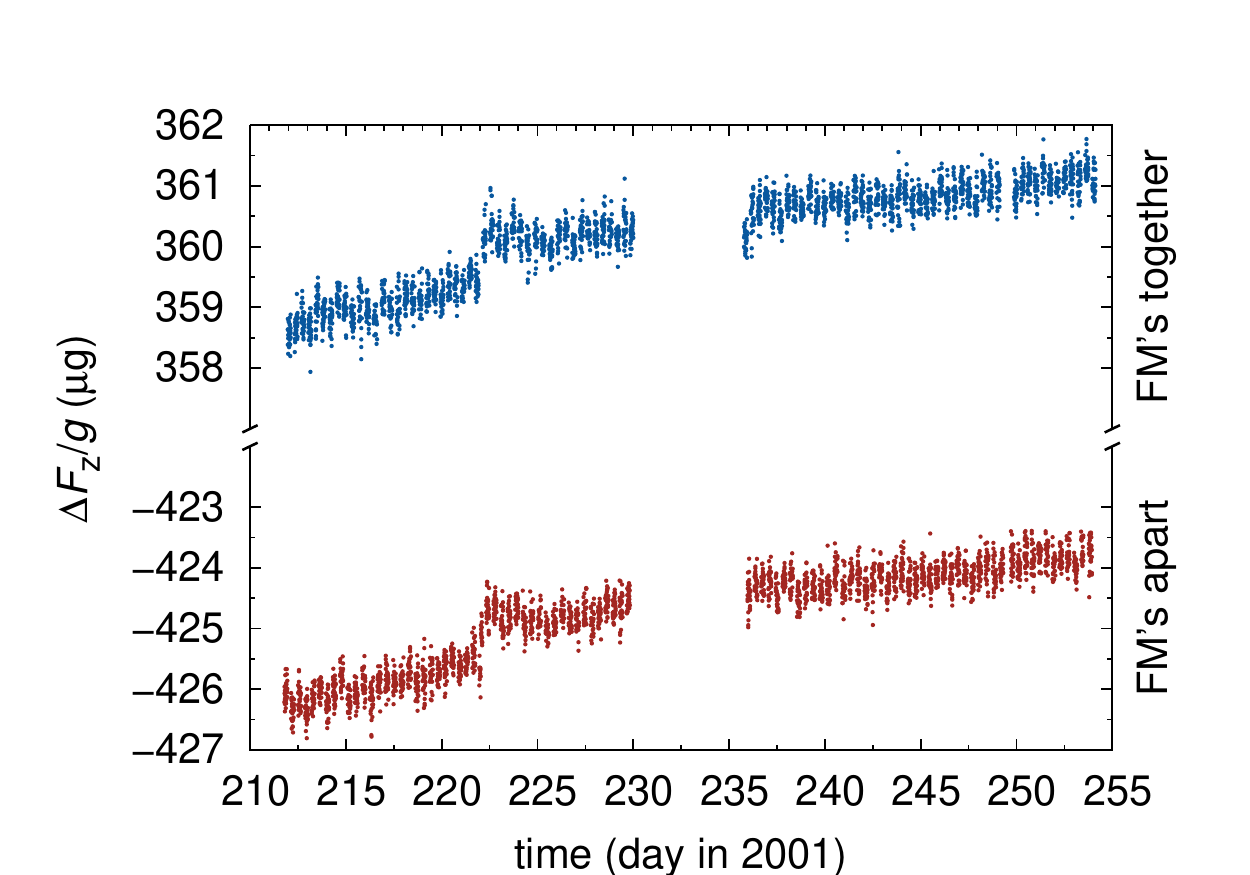}
\includegraphics[width=2.6in]{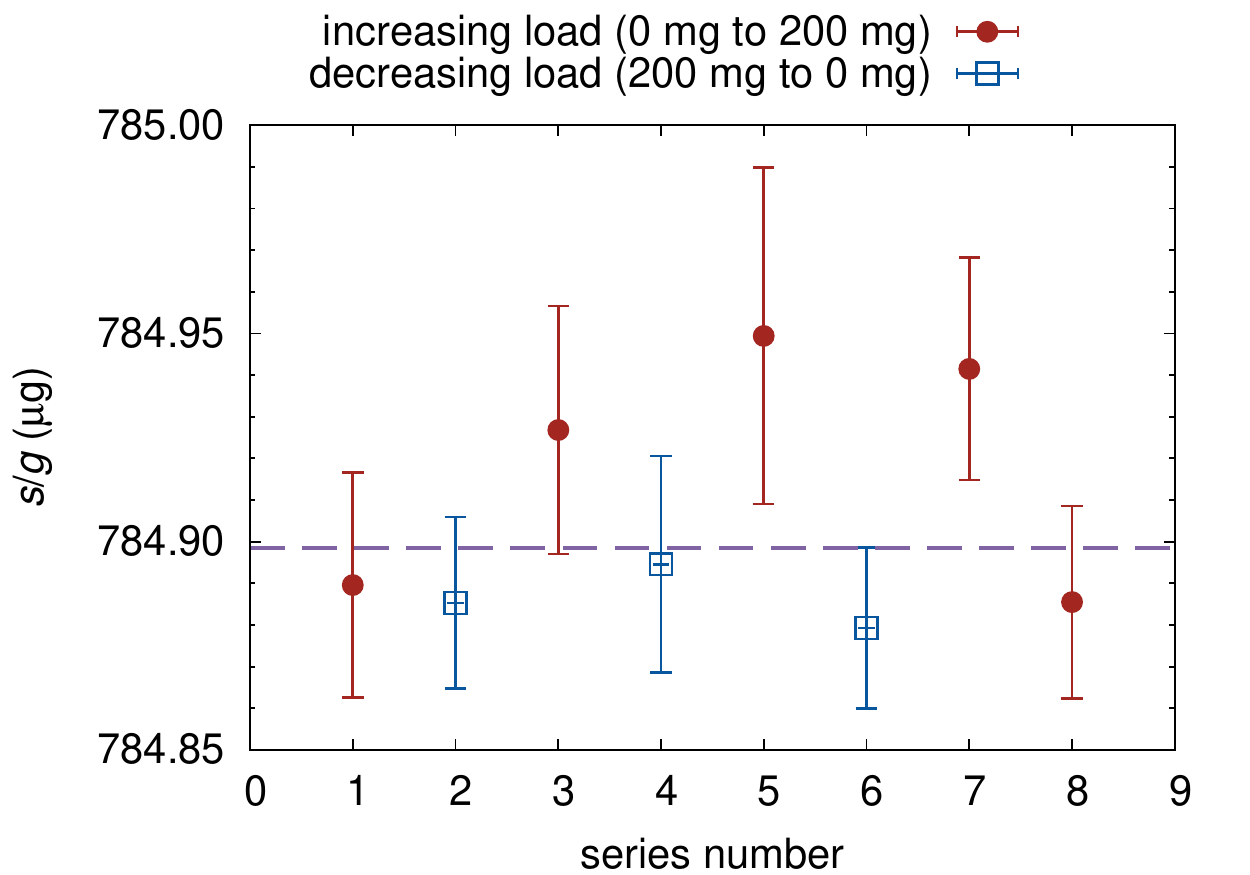}
\end{center}
\caption{On the left, the measured weight difference in $\mu$g
between TM's for the FM's positions apart and together. Note the 
break in the vertical axis. On the
right, the average signal for each of the eight cycles. Cycles
 with increasing load are shown as circles. Cycles with
 decreasing load are shown as squares. The dashed line is the
 average of all eight series. The error bars give the type A
 uncertainties of the weighings.} \label{fig:tmdiff}
\end{figure}

The data used in the final $G$ result were measured during 44 days in the late summer of 2001. Figure~\ref{fig:tmdiff} shows the measured weight difference
between the two test masses for the two field-mass positions. One can see a balance drift and other disturbances, e.g., a jump on day 222 most likely caused
by loss or gain of a dust particle. However, all these effects are common mode, i.e., they cancel in the second difference.

The data shown in figure~\ref{fig:tmdiff} consists of eight nearly complete cycles. A cycle is a measurement of $G$ with all 256 possible load values that
can be achieved with the two sets of auxiliary masses, two test masses and the apart and together positions of the field masses. There are 2304 individual
weighings in one complete cycle. The load was either applied in an increasing (from 0\,mg to 200\,mg) or decreasing (from 200\,mg to 0\,mg) fashion. Of the
eight cycles, five were taken in increasing order. The measured value of the gravitational signal is shown on the right panel of figure~\ref{fig:tmdiff}.

The uncertainty budget is shown in table~\ref{tab:relerrs}. The largest contribution is from the statistical scatter of the weighing data contributing a
relative uncertainty of $11.6\times 10^{-6}$. The next largest effect is water sorption on the test masses. Combining type A and type B uncertainties in
this category yields a relative uncertainty of $10.5\times 10^{-6}$. This effect is due to a different air flow around the vacuum tube for the FMs in
position A and T. Hence, the temperature of the vacuum tube changes slightly; about 0.04\,K and 0.01\,K for the regions around the upper and lower test mass respectively. This temperature change will result in a redistribution of the adsorbed water layers on the vacuum tube and the test masses.
Changing the water layer on the test masses coherently with the field mass position will introduce a bias into the gravitational signal.

The effect of the temperature change was measured in a dummy experiment. The FM's were not moved, but the vacuum tube surrounding the test masses was heated
while the weight variation was recorded. The heating of the vacuum tube produced a temperature variation about 10 times that observed in the $G$ experiment. Scaling the result of the dummy measurement to the experimental conditions yielded the result that the measured signal of $784.899\,4\,\mu$g needed to be corrected by $-0.0168(82)\,\mu$g. This correction resulted in a type A relative standard uncertainty of $7.4\times 10^{-6}$. An additional $7.4\times 10^{-6}$ was assigned as a type B relative uncertainty to reflect the uncertainty in scaling the temperature conditions to those of the actual $G$ measurement. This result supersedes the $49\times 10^{-6}$ relative uncertainty listed in the 1998 experiment. That value was an estimate based on a measurement in which the temperature change of the vacuum tube near the upper test mass was approximately 200 times the amplitude of the temperature variation in the $G$ experiment.

\begin{table}
\caption{Relative type A and type B uncertainties of $G$ as determined in this experiment. All numbers are relative standard uncertainties $(k=1)$.}
\label{tab:relerrs}
\begin{center}
\begin{tabular}{l r r}
{Description}& \multicolumn{1}{c}{Type A} & \multicolumn{1}{c}{Type B}\\
 &\multicolumn{1}{c}{($10^{-6}$)} &\multicolumn{1}{c}{($10^{-6}$)}
\\ \hline
Weighings & 11.6 & \\
TM-sorption  & 7.4 & 7.4\\
Linearity  &6.1& \\
Calibration & 4.0 & 0.5\\
Mass Integration & 5.0&3.3\\
\hline
Total &  16.3 & 8.1 \\
\end{tabular}
\end{center}
\end{table}

The final result of the Z\"{u}rich $G$ experiment is
\begin{equation}
G=\Gres\,\Gunit.
\end{equation}

\begin{figure}[ht!]
\begin{center}
\includegraphics[width=5.8in]{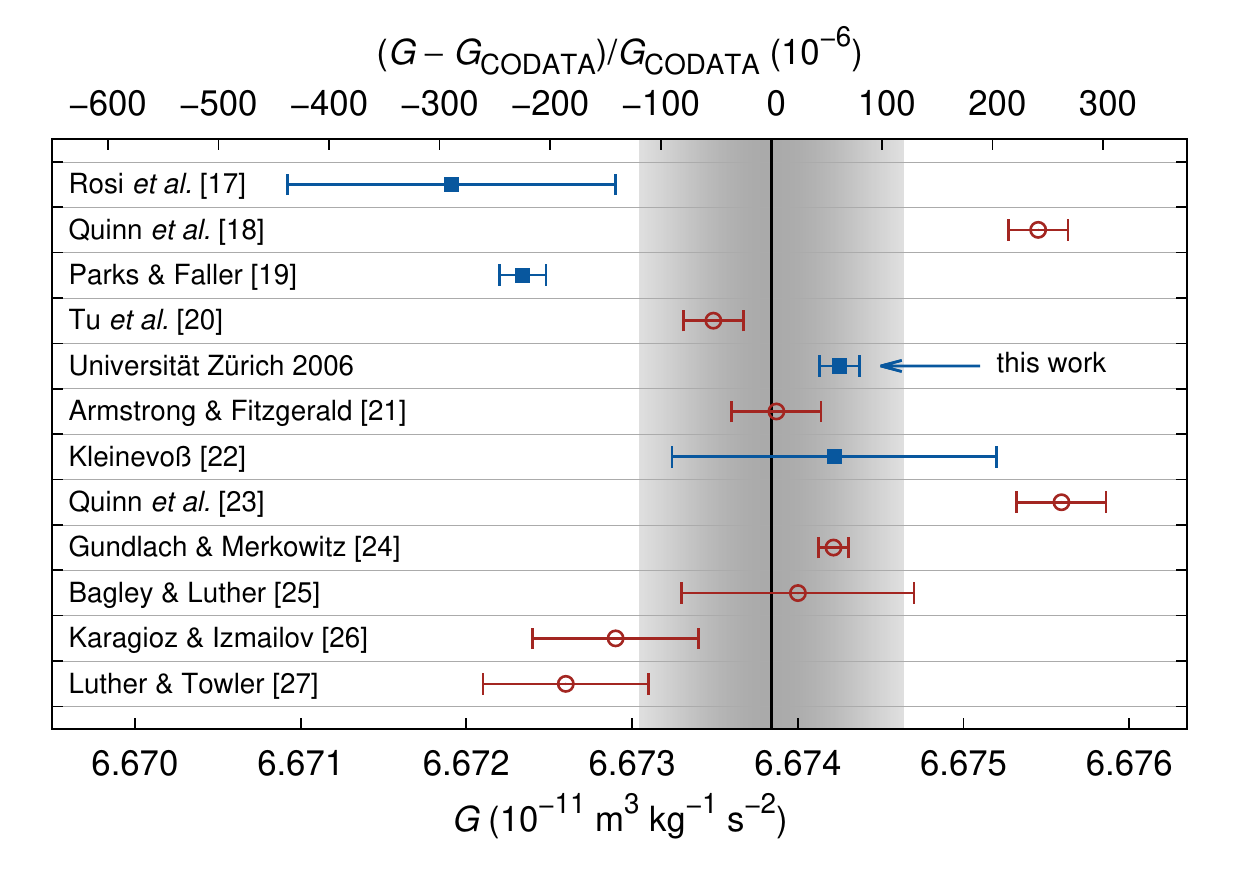}
\end{center}
\caption{The result of the Z\"{u}rich $G$ experiment in relation to the results of other experiments. The experiments denoted by open circles were using
 torsion balances to measure $G$. The four data points denoted solid squares were obtained by different methods. One measurement was performed using an
atom interferometer, two measurements with a pair of pendula, and ours utilized a beam balance. The shaded area in the center of the graph denotes the
$1-\sigma$ confidence interval of the CODATA adjusted value from the 2010 adjustment~\cite{CODATA10}. The years indicate when the results were published.
The values and uncertainties can be found in the references \cite{ Rosi14,Quinn13,Parks10,
Tu10,Armstrong03,Kleinevoss02,Quinn01,Gundlach00, Bagley97,Karagioz96,Luther82}.} \label{fig:Gcomp}
\end{figure}

Figure~\ref{fig:Gcomp} shows this result in relationship to other results. It is noteworthy that the Z\"{u}rich $G$ experiment is only one of four
experiments that do not use torsion balances. Of the three other experiments, two employed simple pendula and one experiment used an atom interferometer.
Out of these four experiments only two, the Z\"urich experiment and the experiment by Parks and Faller~\cite{Parks10} have reached a relative standard uncertainty below $10^{-4}$.

\section{Discussion of the experiment}

Almost a decade has past since the final report of the Z\"{u}rich $G$ experiment was published. In this decade, our experience and views evolved and in
hindsight we would like to give an honest assessment of certain features of the experiment. From this vantage point, different aspects seem more important
than in the past when all of us were immersed in performing or analyzing the experiment.

\subsection{Calibration}
The strongest point of the experiment is that it has a traceable and credible calibration. In section~\ref{sec:cal} most details on  the calibration of the
balance are given. The calibration was performed in situ. Furthermore, unlike in most torsion balance experiments, the calibration was performed in the
same mode, i.e., the configuration of the experiment remained the same. Another interesting aspect of this experiment is the fact that the calibration takes advantage of gravitation itself. The gravitational force between a known mass and the earth was used. This is in contrast to torsion balances, which are often calibrated using electrostatic forces.

\subsection{Nonlinearity}
In 1995 Kuroda pointed out~\cite{Kuroda95} a nonlinear effect in torsion balances that could lead to a significant systematic bias in the so-called time of
swing method. This nonlinearity arises from anelasticity~\cite{Speake99} in the torsion fiber, i.e., the torsional spring constant is a function of frequency and can add a relative systematic bias up to $10^{-4}$ to these experiments.
With Kuroda's publication nonlinearity became a prime concern in experiments determining the gravitational constant. Unfortunately, most of the
effort is spent on anelasticity of torsion fiber. This is certainly an important, but not the only source, of nonlinearity.

Nonlinearity can occur in any instrument. Linearity of an experimental apparatus should never be taken for granted. After obtaining a first result with the
Z\"urich $G$ experiment in 1998, the experimenters were unable to ascertain the linearity of the mass comparator and a generous relative uncertainty of
$200\times 10^{-6}$ was assigned to it.

Subsequently the nonlinearity was investigated, see section~\ref{sec:cal}. The result showed a remarkably small contribution due to nonlinearity of the
mass comparator. The relative standard uncertainty due to nonlinearity was only $6.1\times 10^{-6}$. Although the nonlinearity of the balance was not a
large effect in our experiment, it was a good investment of time and effort to know this for certain.

\subsection{Large gravitational signal}
The gravitational signal in this experiment is large, i.e., the second difference corresponds to $7.7\,\mu$N. For comparison with the torsion-pendulum measurements of  \cite{Quinn13} and \cite{Gundlach00}, the corresponding gravitational forces are approximately  $0.3\,\mu$N and $1.5\times 10^{-4}\,\mu$N, respectively. Thus, the force producing the signals in these two measurements are a factor of  25 and 50,000 times smaller than that of the Z\"urich experiment.

Having a large gravitational force acting on the test masses reduces the relative size of parasitic forces that arise due to surface potentials, gas
pressure forces, and other disturbances.

\subsection{Symmetry and geometry}
The experiment has beautiful symmetry. This symmetry is one of the key points enabling the precision that the experiment finally achieved. The center of
mass of the field masses remains at one point. The field masses are supported by three spindles that have a right-hand thread on the upper part and a left-hand thread on the lower part. Thus, as one field mass moves down it pushes the other field mass up. Other than friction, no mechanical work is performed
in moving from position A to position T. Therefore, a relatively small motor was used to move the field masses. As an example, to overcome the potential
energy of raising one field mass in the five minutes that it takes to change from position A to T, 180\,W of power would be required.

Besides motor power vs. temperature, the symmetric setup has another advantage: The system does
not tilt. The support structure that connects the field-mass assembly to the ground of the experiment has to take up the same force and torques independent
of the field mass position.

The gravitational field on axis of a hollow cylinder centered  at the origin with inner diameter $R_1$, outer diameter $R_2$, height $2B$, and density
$\rho$ is given by
\begin{equation}
V^{(1)} = g_z(0,0,z)  = 2 \pi \rho G \left(r_2^+ - r_2^- + r_1^- - r_1^+ \right)\;  \mbox{with}\; r_{1,2}^{\pm} = \sqrt{R_{1,2}^2 + \left( z \pm B    \right)^2}.
\end{equation}
The potential off axis can be found easily by using a Taylor expansion
of $g_z(r,0,z)$ for
small $r$. Assuming azimuthal symmetry will yield non-zero coefficients
only for even  exponents of $r$.

From the above equation, it can be seen that the  vertical component of the gravitational field near the end of the hollow cylinder ($r=0, z\approx B$) has
the shape of a saddle  surface, i.e., it has a maximum along the axial direction and a minimum in the radial direction. Since the test masses are located at
the saddle points, the accuracy required to measure their position can be achieved with only moderate effort. The vertical and horizontal positions of both
test masses relative to the field masses were obtained with an accuracy of about 0.1\,mm in the horizontal direction and 0.035\,mm in the vertical direction. The measurements were performed using surveyors' tools: a
theodolite for horizontal positions and a leveling instrument for vertical positions.

While we only had to measure the positions of the test masses with moderate precision, the mass distribution of the inner parts of the vessels had to be
known with high accuracy. Hence, these parts were machined with tight tolerances and measured on a coordinate measuring machine with small uncertainties.
For example, the central bores of the hollow cylinders were honed to achieve the required standard uncertainty of 1\,$\mu$m on the inner diameter.

The uncertainties of the dimensions of the test masses, of the dimensions of the field masses, and of their relative positions sum up to relative standard uncertainty in the final result of $4.95\times 10^{-6}$. This number includes the mercury density and mass via the constraint fit, see~\ref{sec:const}. This uncertainty is part of the $5.0\times 10^{-6}$ listed under Type A and mass integration. The remainder is due to uncertainties in the mass measurements of the steel components of the vessels.

\subsection{Vertical system}
From an objective viewpoint, measuring along the direction of local gravity is a terrible thing to do. One has to measure the minuscule gravitational
signal (785\,$\mu$g) on top of a large background (1\,kg). The ratio of background to signal is $1.3\times 10^6$. Since the sensitivity of the experiment
is $10^{-5}$ times the signal,  the ratio of sensitivity to background  is $1.3\times 10^{11}$. This is a really large ratio for a mechanical experiment.
From these numbers it is obvious that a vertical experiment needs a much higher ratio of sensitivity to background than a horizontal experiment.
 This is one of the reasons why several attempts in this geometry achieved
 uncertainties of $10^{-3}$ or larger~\cite{Schwarz99,Lamporesi08}.

Two factors mediate this disadvantage in the Z\"{u}rich $G$ experiment: First, the mass comparator is constructed such that a large fraction of the applied
force is balanced by a built in counter mass and only a small part of the force ($4\times 10^{-3}$) has to be produced by an electromagnetic actuator.
Second, state of the art mass comparators have a fantastic sensitivity. The commercial version of the AT1006 has a relative resolution of $10^{-10}$.

\subsection{Liquid field masses}
\label{sec:const}
The experiment employs two liquid field masses, i.e., 90\% of the gravitational signal is contributed by liquid mercury and only about 10\% by both
vessels. A liquid minimizes density inhomogeneities at the expense of requiring a vessel that can hold the liquid. Indeed, the uncertainty in the density
inhomogeneity of the mercury was only a small contribution to the total uncertainty of the experiment. The absolute value of the density of mercury was
determined at the Physikalisch Technischen Bundesanstalt (PTB) in Germany with a relative uncertainty of $3\times 10^{-6}$. From this measurement, the
known temperature, thermal expansion and isothermal compressibility a density profile of the mercury in the vessels could be calculated.  Modeling the
vessels for the mass integration was cumbersome. Both vessels were broken up in 1200 objects, with positive and negative density. Simple rotational shapes
with rectangular, triangular or circular cross sections were used. Negative density was employed to take away from previously modeled space. This was used,
e.g., for bolt holes, and O-ring grooves.

While mercury had a known density gradient due to its compressibility, density variations in the stainless steel used to manufacture the vessels were unknown. This was especially
problematic for the parts that were close to the test masses.
A density variation in these parts could have affected the result of the experiment. In order to investigate this,
the inner pieces of the vessels were cut into rings after the experiment ended (three rings near the top and three near the bottom of each central tube). The density of each ring was determined by weighing each ring and measuring its dimensions with a coordinate measuring machine.  With this method,  a standard uncertainty of 0.001\,g\,cm$^{-3}$ was achieved. The average density of the rings  was 7.908\,g\,cm$^{-3}$ . The largest variation of the density  was only 0.1\,\%,  which had only a negligible effect on the mass-integration constant.

Another interesting issue arose from deformation after loading the vessels with mercury. The four main parts used to assemble each vessel were carefully
measured on a coordinate measuring machine (CMM). However, the load and hydrostatic pressure deformed the vessels. To take this into account, the shape of
the top and bottom of the full vessels were measured with a laser tracker (LT). From the CMM and LT  measurements combined with equations describing the
bending of thin cylindrical shells~\cite{Timo87} the final shape of the vessels could be determined.

The liquid field masses had one, originally not anticipated, advantage: Since the geometrical dimensions of one mercury volume (inner radius $r$, outer radius
$R$, height $h$),  the mass of the mercury $m$, and the density of the mercury was known, one could set up a system of equations with a constraint on the
volume, mass and density. By using a least squares adjustment of $\chi^2$ given by
\begin{equation}
\chi^2=\left(\frac{r-r_0}{\sigma_r}\right)^2
+\left(\frac{R-R_0}{\sigma_R}\right)^2
+\left(\frac{h-h_0}{\sigma_h}\right)^2
+\left(\frac{m-m_0}{\sigma_m}\right)^2
+\left(\frac{\rho-\rho_0}{\sigma_\rho}\right)^2
\end{equation}
the uncertainties on the four parameters can be substantially improved. Here, the quantities with subscript denote measured values or their standard uncertainties and  those without subscript refer to the adjusted values. This procedure introduces correlations between the adjusted
parameters, but this is a small price to pay compared to the improvement in uncertainty: By using the constraint the uncertainty component caused by the
geometric uncertainties in the mass integration could be reduced by a factor of 7.

\section{Summary}
From 1994 to 2006 seven scientists worked on the Z\"{u}rich $G$ experiment with various degrees of overlap. Their combined work resulted in the value
$G=\Gres\,\Gunit$. The relative standard uncertainty of the measurement is $18\times 10^{-6}$. In recent history this experiment is only one of two
experiments that have not used a torsion balance and have produced a result with a relative uncertainty smaller than $100\times 10^{-6}$.

Torsion balances are very sensitive instruments and are well adapted to the measurement of small forces such as gravitational forces. However, calibrating
a torsion balance is difficult and error-prone~\cite{Michaelis04}. Furthermore, nonlinearities can be a significant effect in measurements using torsion
balances~\cite{Kuroda95}.  Although the calibration of a beam balance is simple and robust, the gravitational signal has to be measured in the presence of
a very large background due to the weight of the test mass. However, as we have demonstrated, it is possible to obtain  a relative statistical accuracy of
several parts in  $10^6$ with this method.

Considering the differences and difficulties of the various approaches, only more data will finally help to resolve the debate concerning the true value of
$G$. We hope to see many different experimental approaches in the future and we encourage the researchers to invest in a credible and traceable calibration
scheme.

%%%%%%%%%% Insert bibliography here %%%%%%%%%%%%%%

\end{document}